\documentclass[useAMS,usenatbib]{mn2e}
\usepackage[totalwidth=515pt,totalheight=660pt,left=1.4cm,right=1.4cm]{geometry}
\usepackage{graphicx,amssymb,color,amsmath}
\usepackage[normalem]{ulem}
\input psfig

\title[GR-induced orbital drift at the pericentre]
{A simple bound for the variation at closest approach of a small body and star due to general relativity}
\author[Veras]
{Dimitri Veras$^{1}$\thanks{E-mail:d.veras@warwick.ac.uk}
\\
$^{1}$Department of Physics, University of Warwick, Coventry CV4 7AL, UK
}

\begin{document}

\date{Accepted 2014 April 07. Received 2014 March 26; in original form 2014 February 23}

\pagerange{\pageref{firstpage}--\pageref{lastpage}} \pubyear{XXXX} 

\maketitle

\label{firstpage}

\begin{abstract}
As a comet, asteroid or planet approaches its parent star, the orbit changes shape due to the curvature of spacetime. For comets in particular, the deviation at the pericentre may noticeably change their ephemerides and affect the dynamics of outgassing, tidal disruption or other processes which act on orbital timescales and are assumed to follow Newtonian gravity. By obtaining and analysing the unaveraged equations of motion in orbital elements due to the dominant post-Newtonian contribution (1PN), I derive a simple analytic expression for the maximum deviation in terms of only the stellar mass and eccentricity of the orbit. This relation can be used to assess the potential importance of including short-period relativistic terms in models containing comets, asteroids or planets, and help determine the level of precision needed in numerical integrations. The magnitude of the deviation in systems with Solar-like stars is typically comparable to the size of comet nuclei, and the direction of the deviation is determined by the eccentricity. I show that for eccentricities above a critical value of $\sqrt{19} - 4 \approx 0.359$, the direction is away from the star.
\end{abstract}

\begin{keywords}
comets: general -- Oort Cloud -- minor planets, asteroids: general -- 
planets and satellites: dynamical evolution and stability 
-- planet-star interactions
-- Celestial Mechanics
\end{keywords}

\section{Introduction and Main Result}

The November 2013 perihelion passage and disintegration of comet C/2012 S1 (ISON) \citep{knietal2013} has 
reinvigorated interest about the physical processes comets experience at closest approach to their parent 
stars.  Both sublimation and tidal forces affect the orbit, ephemeris, and the prospect of the comet 
surviving the close encounter intact. As suggested by \cite{maqetal2012}, another potentially important 
effect arises from general relativity (GR), which they added into their model.

That study is not alone. \cite{shayeo1994} reported that for orbits of comets and asteroids, 
incorporating GR can significantly improve orbital solutions. Consequently, investigators 
of comets such as 55P/Tempel-Tuttle have heeded this advice \citep{yeoetal1996}. Also, the update to the 
Marshall Space Flight Center Meteoroid Stream Model \citep{moscoo2008} featured the inclusion of GR.

Linking the metrics of GR to the idea of a force in Newtonian gravity can be challenging, 
but is well-elucidated in the Appendix of \cite{bengal2008}.  That paper also presents the 
Einstein-Infeld-Hoffman equation \citep{einetal1938}, which provides the corrections to the Newtonian equation of motions 
in a system where every object causes the curvature of spacetime. Cometary studies may assume that only the 
parent star causes such a perturbation because this approximation is excellent, given the many orders of difference in 
mass between a comet and a star. This approximation, which is used in this paper, is also suitable for 
asteroids orbiting stars, planets orbiting stars, and even comets which suffer close encounters with planets, 
such as Shoemaker-Levy 9 did with Jupiter.

The curvature of spacetime due to the star will cause the comet's orbit to deviate from a perfect ellipse, 
parabola or hyperbola.  In the bound (elliptic) case, this curvature causes the long-term (or secular) precession 
of the argument of pericentre, a well-known 
effect for the planet Mercury\footnote{When referring to the word {\it pericentre} by itself, I indicate the actual osculating closest 
approach distance, and not the longitude of pericentre nor argument of pericentre. Some relativity-based 
studies use the former as shorthand for the latter.}. The comet's osculating semimajor axis and eccentricity 
do not change over long timescales, but do change during a single orbit. The magnitude of the change depends 
on both the orbital and spin properties of the comet, which enter into the equations of motion at different 
orders of powers of the speed of light \citep[see equation 3.1a of][]{buoetal2013}. The leading order is independent of spin and is known as the 1PN term. 

Here I isolate and quantify the effect of the 1PN term on a single cometary close passage 
to the star. My main result, which is derived in the next section, is 

\begin{eqnarray}
\Delta_{\rm max} &\approx& \frac{2 G M_{\star}}{c^2} 
\frac{\left(e^2 + 8 e - 3\right)}{\left(1 + e\right)^2}
\nonumber
\\
&\approx&
2.95 \ {\rm km}
\left( \frac{M_{\star}}{M_{\odot}} \right)
\frac{\left(e^2 + 8 e - 3\right)}{\left(1 + e\right)^2}
\label{main}
\end{eqnarray}

\begin{figure}
\centerline{
\psfig{figure=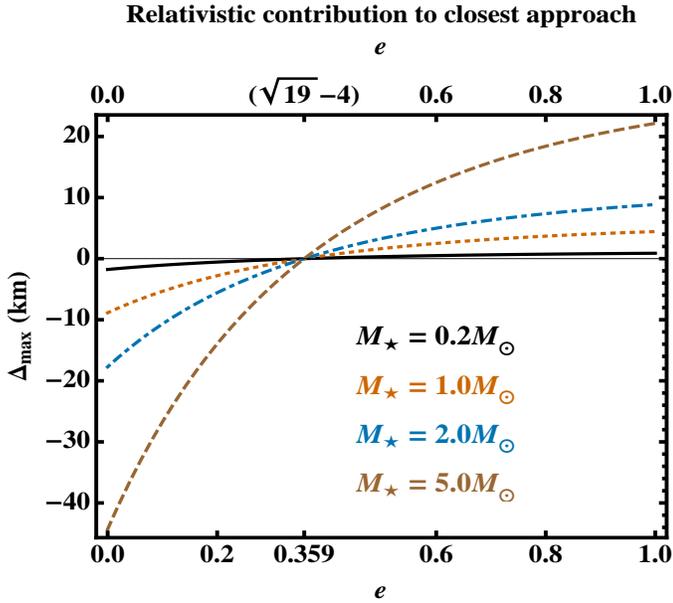,height=8.0cm,width=9.3cm}
}
\caption{
How the orbital pericentre changes when effects from general relativity are included. 
The solid, short-dashed, dot-dashed, and long-dashed lines correspond to stellar
masses of $0.2M_{\odot}$, $1.0M_{\odot}$, $2.0M_{\odot}$, and $5.0M_{\odot}$. 
This plot is based on equation (\ref{main}), and illustrates how the change is
comparable to comet nuclei sizes (of km) for Solar-mass stars. For highly
eccentric bodies, such as observed comets originating from the Oort cloud, relativity pushes the comets away from rather than towards the star at the closest approach. 
}
\label{mainfig}
\end{figure}

\noindent{where} $e$ is the orbital eccentricity, $M_{\star}$ is the stellar mass, and $c$ is the speed of light. The quantity $\Delta$ represents the actual
closest encounter distance (including relativity) minus the closest encounter distance
predicted by Newtonian gravity alone.  For a nearly-parabolic cometary
orbit in a system with a Solar-mass star, GR would increase
the Newtonian pericentre by about 4.4 km.  Figure \ref{mainfig} graphically illustrates the dependence of
the deviation on eccentricity and stellar mass.

This result illustrates that typical deviations are a few km, which is comparable
to typical sizes of cometary nuclei \citep{donrah1982}. Close-up imaging by spacecraft
has helped to establish these sizes \citep[e.g.][]{keletal2013}. Also, Hubble Space Telescope Wide Field 
Camera 3 images suggest that the pre-disruption radius of the nucleus of ISON was no larger 
than 2 km \citep{keletal2014}.  

The remainder of this Letter includes the derivation of equation (\ref{main}) plus a
description of the unaveraged 2-body 1PN problem [Section 2], 
a short discussion [Section 3] and a summary [Section 4].

\section{Derivation}

\subsection{Equations of motion}

I start by deriving the complete equations of motion in orbital elements for the relative 
orbit of the star and comet in the perturbed two-problem with the 1PN term 
from equation (3.1c) of \cite{buoetal2013}. The orbital elements I seek, in addition to
$e$, are the semimajor axis $a$, inclination $i$, longitude of ascending node $\Omega$,
argument of pericentre $\omega$ and true anomaly $f$. Further, I will specifically compute
changes in the pericentre $q$, apocentre $Q$ and mean motion $n$. Let the mass of the comet be $M_{\rm c}$ and the position and velocity vectors of the orbit
be $\vec{r} = \left(x,y,z\right)$ and $\vec{v} = \left(u,v,w\right)$. Then

\begin{eqnarray}
\frac{d^2\vec{r}}{dt^2} 
&=& 
- \frac{G \left(M_{\star} + M_{\rm c} \right) \vec{r}}{r^3}
+
\delta
\label{eqmotion}
\\
&&
\nonumber
\end{eqnarray}

\noindent{with}

\begin{eqnarray}
\delta &=& 
\frac{2 G  
\left(2M_{\star}^2 + 3M_{\star}M_{\rm c} + 2M_{\rm c}^2\right)}{\left(M_{\star} + M_{\rm c}\right)}
\frac{\dot{r}\vec{v}}{c^2 r^2} 
\nonumber
\\
&\ & \ - 
\frac{G  
\left(M_{\star}^2 + 5M_{\star}M_{\rm c} + M_{\rm c}^2\right)}{\left(M_{\star} + M_{\rm c}\right)}
\frac{\vec{r} v^2}{c^2 r^3} 
\nonumber
\\
&\ & \ +
\frac{3 G M_{\star}M_{\rm c}}{2\left(M_{\star} + M_{\rm c}\right)}
\frac{\vec{r} \dot{r}^2}{c^2 r^3} 
\nonumber
\\
&\ & \ +
2 G^2 \left(2M_{\star} + M_{\rm c}\right)\left(M_{\star} + 2 M_{\rm c}\right)
\frac{\vec{r}}{c^2 r^4}
,
\label{GRpert}
\end{eqnarray}

\noindent{where} the overdot refers to a time derivative 
so that 

\begin{eqnarray}
\dot{r} &=& \frac{ux + vy + wz}{\sqrt{x^2 + y^2 + z^2}} = \frac{\vec{r}\cdot\vec{v}}{r}
.
\\
&&
\nonumber
\end{eqnarray}

By expressing equation (\ref{GRpert}) in terms
of Cartesian positions and velocities, I can apply the formalism of \cite{vereva2013a} to generate
the unaveraged equations of motion in orbital elements. This technique has now been applied
to the perturbed two-body problem with a variety of forces, including Galactic tides \citep{vereva2013b},
anisotropic stellar mass loss \citep{veretal2013} and additional bodies in the restricted 
$N$-body problem \citep{veras2014}.  I obtain

\begin{eqnarray}
\frac{da}{dt} &=&
\frac{G^2 e \left(1 + e \cos{f}\right)^2 \sin{f}}{2 c^2 a^3 n \left(1 - e^2\right)^{7/2}} 
\nonumber
\\
&\ & \ \times
\bigg[
4 \left(7 + 3 e^2\right) (M_{\star}^2 + M_{\rm c}^2) + \left(44 + 7 e^2\right) M_{\star} M_{\rm c} 
\nonumber
\\
&\ & \  +
8 e \left(5M_{\star}^2 + 6 M_{\star} M_{\rm c} + 5M_{\rm c}^2\right) \cos{f}
\nonumber
\\
&\ & \  -
3 e^2 M_{\star} M_{\rm c} \cos{\left(2f\right)}
\bigg]
,
\label{GRdadt}
\\
\frac{dn}{dt} &=&
-\frac{3}{2} \frac{n}{a} \frac{da}{dt}
,
\label{GRdndt}
\end{eqnarray}

\begin{eqnarray}
\frac{de}{dt} &=&
\frac{G^2 \left(1 + e \cos{f}\right)^2\sin{f}}{4 c^2 a^4 n \left(1 - e^2\right)^{5/2}} 
\nonumber
\\
&\ & \ \times
\bigg[
4 \left(3 + 7 e^2\right) (M_{\star}^2 + M_{\rm c}^2) + \left(20 + 31 e^2\right) M_{\star} M_{\rm c} 
\nonumber
\\
&\ & \ +
8 e \left(5M_{\star}^2 + 6 M_{\star} M_{\rm c} + 5M_{\rm c}^2\right) \cos{f}
\nonumber
\\
&\ & \ -
3 e^2 M_{\star} M_{\rm c} \cos{\left(2f\right)}
\bigg]
,
\label{GRdedt}
\end{eqnarray}

\begin{eqnarray}
\frac{di}{dt} &=& 
\frac{d\Omega}{dt} = 0
,
\\
&&
\nonumber
\\
\frac{d\omega}{dt} &=&
\frac{G^2 \left(1 + e \cos{f}\right)^2}{8 c^2 a^4 e n \left(1 - e^2\right)^{5/2}} 
\nonumber
\\
&\ & \ \times
\bigg[
\big[8 \left(e^2 - 3\right) (M_{\star}^2 + M_{\rm c}^2)
\nonumber
\\
&\ & \ +
\left(37 e^2 - 40\right) M_{\star} M_{\rm c} \big]
\cos{f} 
+ 
24 e \left(M_{\star} + M_{\rm c}\right)^2
\nonumber
\\
&\ & \  -
8 e \left(5M_{\star}^2 + 6 M_{\star} M_{\rm c} + 5M_{\rm c}^2 \right) \cos{\left(2f\right)}
\nonumber
\\
&\ & \  +
3 e^2 M_{\star} M_{\rm c} \cos{\left(3f\right)}
\bigg]
,
\label{GRdomdt}
\\
\frac{df}{dt} &=& 
\frac{n \left(1 + e \cos{f}\right)^2}{\left(1 - e^2\right)^{3/2}}
                - \frac{d\omega}{dt} 
,
\label{GRdfdt}
\end{eqnarray}

\noindent{}where $n = \sqrt{G \left(M_{\star} + M_{\rm c}\right)}a^{-3/2}$.  
Note that $d\omega/dt = d\varpi/dt$, where $\varpi$ is the longitude
of pericentre.  This characterisation of the orbital element evolution due to the 1PN term 
is not new, and has been described in various forms by, for example, \cite{brumberg1972,brumberg1991}, and more recently
in pgs. 505-508 of \cite{kopetal2011}, and \cite{li2012}.  In the unperturbed Newtonian two-body problem, none of these orbital element variations exist except for the first term of equation (\ref{GRdfdt}).

Another useful quantity is the time evolution of the comet-star separation.
Through equations
(\ref{GRdadt}), (\ref{GRdedt}) and (\ref{GRdfdt}) and with
$r = a \left(1 - e^2\right)/\left(1 + e \cos{f}\right)$, I find

\begin{eqnarray}
\frac{dr}{dt} &=& 
e \sin{f} \sqrt{\frac{G \left(M_{\star} + M_{\rm c}\right)}{a\left(1 - e^2\right)}}
,
\label{GRdrdt}
\\
&&
\nonumber
\end{eqnarray}

\noindent{}which is equivalent to the unperturbed two-body term.  Therefore, post-Newtonian variations in $dr/dt$ enter indirectly through $a$, $e$ and $f$.

Now consider the time evolution of the pericentre $q = a\left(1 - e\right)$ 
and apocentre $Q = a\left(1 + e\right)$.  From equations 
(\ref{GRdadt}-\ref{GRdedt}), I obtain

\begin{eqnarray}
\frac{dq}{dt} &=& 
\frac{G^2 \left(1 + e \cos{f}\right)^2\sin{f}}{4 c^2 a^3 n \left(1 - e\right)^{3/2} \left(1 + e\right)^{7/2}} 
\nonumber
\\
&\ & \ \times
\bigg[
4 \left(-3 + 8e + e^2\right) (M_{\star}^2 + M_{\rm c}^2) 
\nonumber
\\
&\ & \ + \left(-20 + 48 e + 17 e^2\right) M_{\star} M_{\rm c} 
\nonumber
\\
&\ & \ -
8 e \left(5M_{\star}^2 + 6 M_{\star} M_{\rm c} + 5M_{\rm c}^2\right) \cos{f}
\nonumber
\\
&\ & \ +
3 e^2 M_{\star} M_{\rm c} \cos{\left(2f\right)}
\bigg]
\label{dqdt1}
\end{eqnarray}

\noindent{}and

\begin{eqnarray}
\frac{dQ}{dt} &=& 
\frac{G^2 \left(1 + e \cos{f}\right)^2\sin{f}}{4 c^2 a^3 n \left(1 + e\right)^{3/2} \left(1 - e\right)^{7/2}} 
\nonumber
\\
&\ & \ \times
\bigg[
4 \left(3 + 8e - e^2\right) (M_{\star}^2 + M_{\rm c}^2) 
\nonumber
\\
&\ & \ + \left(20 + 48 e - 17 e^2\right) M_{\star} M_{\rm c} 
\nonumber
\\
&\ & \ +
8 e \left(5M_{\star}^2 + 6 M_{\star} M_{\rm c} + 5M_{\rm c}^2\right) \cos{f}
\nonumber
\\
&\ & \ -
3 e^2 M_{\star} M_{\rm c} \cos{\left(2f\right)}
\bigg]
.
\label{dQdt1}
\end{eqnarray}

\subsection{Properties of equations}

Equations (\ref{GRdadt}-\ref{dQdt1}) shed light on the behaviour of the comet 
in ways that equations (\ref{eqmotion}-\ref{GRpert}) do not. For example, 
the averaged values of 
$a$, $e$, $q$, $Q$, $n$ and $r$ all equal zero, meaning that none of these elements showcase any secular variations.  The averaged value
of $\omega$ gives $3 \left[G \left(M_{\star} + M_{\rm c}\right) \right]^{3/2}/[a^{5/2}c^2 \left(1 - e^2\right)]$, which famously helped explain Mercury's long-term motion.

Additionally, and of interest here, is that the orbital element equations provide physical
intuition for how the motion changes during a single orbit. The stationary
points of equation (\ref{GRdrdt}) are $f = 0^{\circ}$ and $f = 180^{\circ}$, demonstrating
that the comet-star distance monotonically decreases from the maximum separation
to the minimum separation even with the inclusion of GR. This finding is important because the osculating pericentre evolution 
(equation \ref{dqdt1}) deceivingly does not exhibit the same behaviour, containing additional
stationary points.

\begin{figure}
\centerline{
\psfig{figure=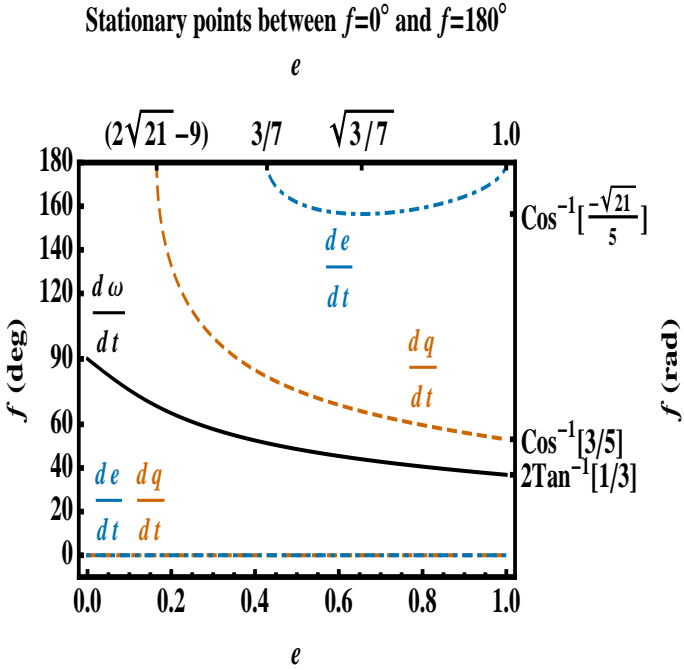,height=9.0cm,width=9cm}
}
\caption{
The stationary points between $f = 0^{\circ}$ and $f = 180^{\circ}$
for equations (\ref{GRdedt}), (\ref{GRdomdt}) and  (\ref{dqdt1}).
These curves demonstrate that the osculating values of $e$, $q$ 
and $\omega$ do not evolve monotonically between the times of 
furthest and closest approach of the comet and star.
}
\label{stationary}
\end{figure}

I now give a summary of all the stationary points, assuming that $M_{\rm c} = 0$.  
This assumption, which I will carry through the remainder of the paper, is excellent
because of the many orders of magnitude difference in the masses of a comet and star.
The variables $a$, $e$, $q$, $Q$,
$n$, and $r$ all become stationary at $f = 0^{\circ}$ and $f = 180^{\circ}$.
The variable $\omega$ does not.  The variables $e$, $q$ and $\omega$ each contain 
2 other stationary points which are symmetric about the major axis of the osculating
ellipse at apocentre.  These stationary points are functions of both the eccentricity
and true anomaly, and the ones between $f = 0^{\circ}$ and $f = 180^{\circ}$ are shown 
graphically in Figure \ref{stationary}. Critical points of these curves are displayed
on the left, upper and right axes, and the curves are described by the following 
explicit functions

\begin{eqnarray}
f_{e}(e) &=& \arccos{\left[\frac{-3-7e^2}{10e}\right]}
,
\\
f_{q}(e) &=& \arccos{\left[\frac{-3 + 8 e + e^2}{10 e}\right]}
,
\\
f_{\omega}(e) &=& 2 \arctan{  
\sqrt{
\frac
{18e - \sqrt{9 + 314 e^2 + e^4}}
{e^2 + 2e - 3}
}   
}
.
\end{eqnarray}

\subsection{Close approach variation}

Having described the properties of the motion, I can
now estimate the GR-induced variation at the closest approach.  Ideally, 
equations (\ref{GRdadt}-\ref{dQdt1}) would be completely 
solvable. However, until a solution is found, another approach is needed.


Define $\Delta$ as the difference in distance at closest approach from the 
Newtonian and post-Newtonian values. Equivalently, for 
an arbitrary point ``p'' on the approach to the pericentre such that
$\pi \le f_{\rm p} < 2 \pi$,

\begin{eqnarray}
\Delta(f_{\rm p}) &\equiv& \underbrace{r\left(f = f_{\rm p}\right) - r\left(f = 2\pi\right)
}_{\rm GR}
\nonumber
\\
&-&
\underbrace{
\left[ \frac{\Pi(2\pi) - \Pi(f_{\rm p})}{n(f_{\rm p})} \right] 
\frac{n(f_{\rm p})}{2\pi - f_{\rm p}}\int_{f_{\rm p}}^{2\pi}  \frac{dr}{dt} \left(\frac{dt}{df}\right)_{\rm trunc} df
}_{\rm Newton}.
\nonumber
\\
&&
\end{eqnarray}

\noindent{}Here, $(dt/df)_{\rm trunc}$ is equal to the inverse of the 
first term of equation (\ref{GRdfdt}), and the mean anomaly $\Pi$ satisfies

\begin{equation}
\Pi\left(f_{\rm p}\right) = \arccos{
\left[   
\frac{e(f_{\rm p}) + \cos{f_{\rm p}}}
{1 + e(f_{\rm p})\cos{f_{\rm p}}}
\right]}
-
\frac{e(f_{\rm p})\sqrt{1 - e(f_{\rm p})^2} \sin{(f_{\rm p})}}
{1 + e(f_{\rm p})\cos{f_{\rm p}}}
.
\end{equation}

As $r$ monotonically decreases from apocentre to pericentre, the maximum
possible deviation is 

\begin{eqnarray}
\Delta_{\rm max} &\equiv& \Delta\left(f_{\rm p} = \pi \right)
\nonumber
\\
&=& r_{\rm max} - r_{\rm min} - 2 a_{\rm ap} e_{\rm ap}
\label{delx1}
\end{eqnarray}

\noindent{}where $a_{\rm ap}$ and $e_{\rm ap}$ are the semimajor axis and 
eccentricity at apocentre.  I have found that equation (\ref{delx1})
can be well-approximated by

\begin{eqnarray}
\Delta_{\rm max}
&\approx&  
\left[ \frac{-\pi}{n} \right] 
\frac{n}{\pi}\int_{\pi}^{2\pi}  \frac{dq}{dt} \left(\frac{dt}{df}\right)_{\rm trunc} df
,
\label{delx2}
\\
&&
\nonumber
\end{eqnarray}

\noindent{}which simplifies to equation (\ref{main}).  
Equation (\ref{main}) demonstrates that the maximum possible radial drift 
outside of and within the predicted Newtonian values
for a Solar-mass star are about 4.43 km and 8.86 km, respectively.
Also, the equation illustrates 
that the comet's actual pericentre exceeds the
Newtonian value for only a subset of $e$ values. 
The bifurcation occurs exactly at

\begin{equation}
e_{\rm crit} = \sqrt{19} - 4 \approx 0.359
.
\label{ecrit}
\end{equation}

One can determine the goodness of the approximation of equation
(\ref{main}) by comparing $\Delta_{\rm max}$ with the
result of numerical simulations. However, numerical integrations of comets on
wide orbits which must correctly track km-scale or m-scale 
variations can require a
high level of precision. Further, the accuracy of these integrations is limited
by the maximum allowable timestep and the output resolution. The dependence of the necessary precision and accuracy on $M_{\star}$, $a_{\rm ap}$ and $e_{\rm ap}$ is nontrivial.

I validated equation (\ref{main}) for a wide range of triplets
($M_{\star}$, $a_{\rm ap}$, $e_{\rm ap}$) by numerically integrating equations
(\ref{GRdadt}), (\ref{GRdedt}), (\ref{GRdomdt}) and (\ref{GRdfdt}) over one entire orbit using 
the software program {\it Mathematica}.  {\it Mathematica} allows the user to specify the number of 
significant digits retained in each intermediate step of the integration,
along with other options. The phase space is too large to explore in its entirety,
and is dependent on integration technique. 

Nevertheless, I now provide some error estimates,
which were obtained by propagating 35 digits of precision in the integrations
and assuming a Solar-mass central star.  For small bodies with Earth-like 
semimajor axes ($\approx 1$ au), $e_{\rm ap}$ values corresponding to 
$(0.99, 0.50, 0.359, 0.01)$ yielded fractional differences equivalent to
$|({\rm equation} \ \ref{delx1} - {\rm equation} \ \ref{main})/({\rm equation} \ \ref{delx1})|$
of $(8 \times 10^{-4}, 2 \times 10^{-6}, 2 \times 10^{-3}, 1 \times 10^{-7})$.  For
small bodies with orbits as wide as the Kuiper belt ($\approx 30$ au),
and $e_{\rm ap} = (0.999, 0.99, 0.50, 0.359, 0.01)$, I obtained fractional
differences of $(3 \times 10^{-3}, 3 \times 10^{-5}, 6 \times 10^{-8}, 
2 \times 10^{-3}, 4 \times 10^{-9})$.  Note that for some of these simulations,
I have purposely used $0.359$ rather than the precise value of $e_{\rm crit}$.

Scattered disc or inner Oort cloud-like
objects with $a \approx 10^3$ au and $e_{\rm ap} = (0.99999, 0.9999, 0.999)$
yield fractional errors of $(5 \times 10^{0}, 8 \times 10^{-3}, 8 \times 10^{-5})$,
whereas for $a \approx 10^4$ au and $e_{\rm ap} = (0.99999, 0.9999, 0.999)$,
the errors are $(6 \times 10^{-2}, 7 \times 10^{-4}, 8 \times 10^{-6})$.
Outer Oort cloud-like distances of $a \approx 10^5$ au and a range of eccentricities
including $e_{\rm ap} = (0.999999, 0.99999, 0.9999, 0.5, 0.359, 0.01)$ yielded
errors of $(1 \times 10^{0}, 3 \times 10^{-3}, 2 \times 10^{-3}, 2 \times 10^{-9}
, 5 \times 10^{-5}, 9 \times 10^{-10})$.  For the highest values of $a_{\rm ap}$ and lowest
pericentre values, either the approximation breaks down or my numerical integrations
have reached their computational limit. The former case makes sense when $a_{\rm ap} \left(1 - e_{\rm ap}\right)$ is fixed as $a_{\rm ap}$ and $e_{\rm ap}$ increase, because of equation (\ref{GRdomdt}): The approximation of equation (\ref{main}) neglects the contribution of $d\omega/dt$, but this quantity becomes increasingly important because when $M_{\rm c} = 0$, then $d\omega/dt \propto a^{-5/2}\left(1-e^2\right)^{-4}$.

\section{Discussion}

Equation (\ref{main}) is applicable to any relatively small secondary, not just comets.  However, for objects such as planets, the variations at the pericentre are 3-5 orders of magnitude smaller than their physical radii.  For hot Jupiters, which orbit close to their stars ($a \lesssim 0.1$ au), the effect generally causes inward drift toward the star because of the planets' relatively circular orbits.  In fact, as of 18 Feb 2014, only 5 out of 292 such planets\footnote{See the Exoplanets Data Explorer at exoplanets.org} have measured eccentricities greater than $e_{\rm crit}$.

The applicability of Equation (\ref{main}) is not limited to the Solar System.  There are tantalising hints of exocomets in systems with A stars \citep{welmon2013,kieetal2014}. Also, over 25 per cent of all white dwarfs are thought to host remnant planetary systems due to the presence of rocky atmospheric pollutants \citep{zucetal2010}. These pollutants might arise from asteroids or comets on highly eccentric orbits, which could disrupt into dusty discs \citep[e.g.][]{faretal2010} or gaseous discs \citep[e.g.][]{gaeetal2008}. Given that about half of all stars in the Galaxy are more massive than a few tenths of a $M_{\odot}$ \citep{offetal2013}, Figure \ref{mainfig} suggests that the maximum variation at pericentre is comparable to the size of a comet nucleus for most stars in the Milky Way.  Nevertheless, in extrasolar systems with a single main sequence star, the importance of short-period GR terms may be limited to highly accurate numerical simulations until observational capabilities improve.

\section{Summary}
I have derived a simple formula (equation \ref{main}) which approximates the maximum pericentre deviation due to general relativity of a small body in terms of only the eccentricity of its orbit and the stellar mass. Comets on bound near-parabolic orbits will drift away from the star at the pericentre by about 4.4 km $\times (M_{\star}/M_{\odot})$ from the Newtonian value predicted at the apocentre. Because this variation is comparable to the size of comet nuclei, detailed orbital evolution models, as well as models of sublimation and disruption, might need to include short-period effects due to relativity. This Letter also characterises the equations of motion and stationary points of the 1PN two-body problem in terms of osculating orbital elements, attributes which may aid future investigations involving small body ephemerides.

\section*{Acknowledgments}

I thank the referee for astute suggestions which have improved this Letter. I also thank Thomas R. Marsh, Despoina K. Skoulidou, and Elizabeth R. Stanway for helpful discussions. This work benefited from support by the European Union through ERC Grant Number 320964.

\label{lastpage}
\end{document}